# Bypassing the resolution limit of diffractive zone plate optics via rotational Fourier ptychography

**Chengfei Guo**[a,b,†], **Shaowei Jiang**[b,†], **Pengming Song**[c], **Zichao Bian**[b], **Tianbo Wang**[b], **Pouria Hoveida**[b], **Xiaopeng Shao**[a,*]

a *Xi'an Key Laboratory of Computational Imaging, Xidian University, Shaanxi, 710071, China*
b *Biomedical Engineering, University of Connecticut, Storrs, CT, 06269, USA*
c *Electrical and Computer Engineering, University of Connecticut, Storrs, CT, 06269, USA*
† *These authors contributed equally to this work*

**Abstract**. Diffractive zone plate optics uses a thin micro-structure pattern to alter the propagation direction of the incoming light wave. It has found important applications in extreme-wavelength imaging where conventional refractive lenses do not exist. The resolution limit of zone plate optics is determined by the smallest width of the outermost zone. In order to improve the achievable resolution, significant efforts have been devoted to the fabrication of very small zone width with ultrahigh placement accuracy. Here, we report the use of a diffractometer setup for bypassing the resolution limit of zone plate optics. In our prototype, we mounted the sample on two rotation stages and used a low-resolution binary zone plate to relay the sample plane to the detector. We then performed both in-plane and out-of-plane sample rotations and captured the corresponding raw images. The captured images were processed using a Fourier ptychographic procedure for resolution improvement. The final achievable resolution of the reported setup is not determined by the smallest width structures of the employed binary zone plate; instead, it is determined by the maximum angle of the out-of-plane rotation. In our experiment, we demonstrated 8-fold resolution improvement using both a resolution target and a titanium dioxide sample. The reported approach may be able to bypass the fabrication challenge of diffractive elements and open up new avenues for microscopy with extreme wavelengths.

## 1. Introduction

A binary zone plate lens consists of a set of radially symmetric rings, known as Fresnel zones, which alternate between opaque and transparent. Light passing through the zone plate lens will be diffracted around the opaque zones and constructively interferes at the desired focus, creating an image there. There are two parameters for the design of zone plate lens: focal distance $f$ and zone number $n$. Once these two parameters are given, the position of the n[th] zone can be calculated using $r_n = \sqrt{nf\lambda}$, where λ is the incident wavelength. The resolution limit of a binary zone plate lens is determined by the feature size of the outermost zone. If we ignore the aberrations of the zone plate and use it in a high-magnification setting, the numerical aperture (NA) can be approximated by λ/(2Δr), where Δr is the width of the outermost (smallest) zone. For imaging applications, we need to reduce the size of Δr to improve the achievable resolution. Using state-of-art e-beam lithography, Δr can be as small as ~10 nm [1, 2] and the achievable spatial resolution is on the same level of Δr. Further reducing the zone width with nanometer placement accuracy is very challenging for current fabrication techniques [3, 4].

Here, we investigate another option for improving the achievable resolution for diffractive zone plate optics. Instead of pushing the size limit of the outermost zone, we employ a diffractometer setup and use the Fourier ptychographic imaging procedure to bypass the resolution limit of the zone plate optics. The Fourier ptychography (FP) approach is an extensively studied computational imaging modality for high-resolution microscopy applications [5-22]. It can be implemented in both coherent and incoherent settings. For coherent FP, the sample is illuminated with angle-varied plane waves [5]. The corresponding captured sample images are used to recover the high-resolution sample image using an iterative process that is similar to the Gerchberg-Saxton algorithm [23-26]. If the employed optics contains some unknown aberrations, it is possible to recover both the high-resolution sample image and the unknown aberrations at the same time [27, 28]. The final achievable resolution of coherent FP is determined by the largest incident angle of the plane wave illumination, not the numerical aperture (NA) of the employed optics. For incoherent FP, the sample is illuminated with different intensity patterns in an incoherent imaging setting [19, 20]. Similar to the aberration recovery process in coherent FP, it is possible to recover both the unknown intensity patterns and the super-resolution sample image at the same time. The final achievable resolution of incoherent FP is determined by the feature size of the intensity patterns.

The advantages of the FP approach may enable new research directions for microscopy imaging with extreme wavelengths, where refractive lenses do not exist [29-31]. In this case, diffractive optical elements such as binary zone plate are used in the image formation process. While the FP approach has been successfully demonstrated for conventional refractive lenses, there is no guarantee that it will also work for diffractive optical elements that rely on coherent addition of incoming waves in the image formation process. One would further expect that the zero order and other unwanted higher diffraction orders would compound any attempted image recovery. To handle this, we need to exclude the captured images with strong zero-order diffraction in the recovery process. Furthermore, beam stability is an important concern for light sources with extreme wavelengths. It is difficult to change incident angle due to instrumental constraints. Therefore, a new experimental design is needed to ensure that the beam condition does not change during the image acquisition process.

The goal of this paper is to demonstrate, for the first time, the use of a diffractometer setup in conjunction with FP for bypassing the resolution limit of diffractive optical element. The reported imaging procedure may open up new avenues for microscopy with extreme wavelengths, where refractive lenses do not exist. This paper is structured as follows: in section 2, we will report our prototype setup and the operating principle. In section 3, we will demonstrate the imaging performance using a resolution target and a titanium dioxide sample. Finally, we will summarize the results and discuss the future directions in section 4.

## 2. Experimental setup

The schematic of the experimental setup is shown in Fig. 1(a). A binary zone plate is used to relay the sample plane to the CCD detector. The designed binary zone plate was fabricated on a 0.06 inch thick soda-lime glass metallized with a chrome layer and coated with AZ1518 photoresist. A laser pattern generator (Heidelberg, DWL2000) fitted with a 405 nm diode laser was used for writing the zone plate pattern, followed by standard development and chrome etching processes. The remaining AZ1518 photoresist was then completely removed from the photomask before use in the optical setup. The size of the zone plate is about 4 mm by 4 mm (Fig. 1(b1)), and the designed NA of the binary zone plate is about 0.02. The CAD design file of the binary zone plate can be downloaded from Supplementary 1.

In our prototype setup, a collimated beam from a LED (625nm wavelength, M625L3-C1, Thorlabs) is employed for sample illumination, which is defined by a diaphragm with a diameter of 6 mm. The beam conditions remain unchanged during the entire image acquisition process. In order to change the incident angle of the beam, we mounted the sample, the binary zone plate lens, and the CCD detector (DMK 31BU03, The Imaging Source) on a large stage (URS100BCC, Newport) for out-of-plane rotation, similar to a diffractometer setup. We also utilized a second rotation stage (PRM1Z8, Thorlabs) to perform in-plane rotation of the sample. Fig. 1(b1) and (b2) show the in-plane and out-of-plane rotation axes of our prototype setup. Both the sample and the zone plate are mounted on the manual translation stages (PT1, Thorlabs) to allow for three-dimensional adjustments. The whole experimental setup is covered under a black curtain to avoid the noise beam from outside.

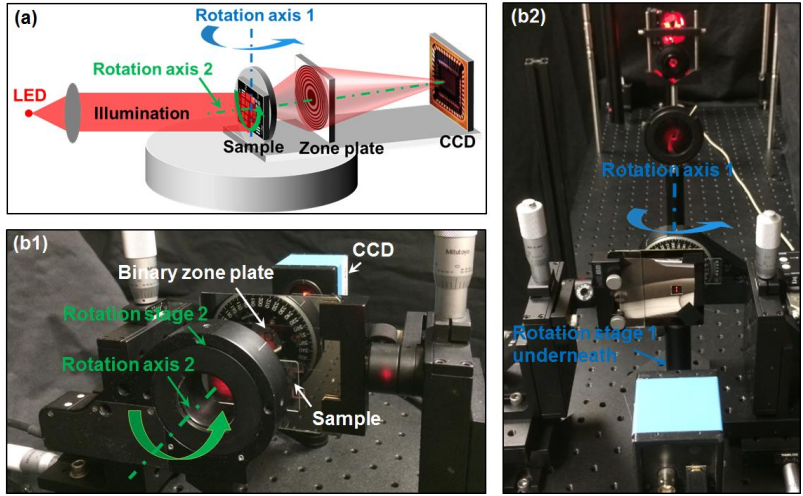

Fig. 1 (a) The experimental setup using two rotation stages. A collimated LED beam was used for sample illumination. The beam conditions remain unchanged during the acquisition process. (b) The sample, the zone plate lens and the CCD detector were mounted on a large rotation stage (axis 1) for out-of-plane rotation. A second stage was used to perform in-plane rotation along axis 2.

The operation of the reported platform can be described as follows: 1) we first perform out-of-plane rotation along axis 1 and capture raw sample images corresponding to different incident angles. 2) The captured images will be used to recover a high-resolution sample image using a modified Fourier ptychographic procedure. The resolution improvement is along the tangent direction of stage 1. 3) We will then rotate the sample to another in-plane orientation using stage 2 and repeat step 1 and 2. By doing so, we can improve resolution of the final image along other directions (we will discuss the detailed implementation in the next section).

Figure 2(a1)-(a4) illustrate the schematic when performing the out-of-plane rotation with four different angles. In Fig. 2(b1)-(b4), we show the captured raw images of a resolution target with the range of out-of-plane rotation angles from 0 degree to 9 degrees, and the in-plane rotation angle is set to 0 degree. While in Fig. 2(c1)-(c4), the in-plane rotation angle is set to 90 degrees, and the corresponding captured raw images of the resolution target are presented. By comparing Fig. 2(b1) and 2(c1), one can see that the orientation of the sample is rotated by 90 degrees with the in-plane rotation of axis-2.

## 3. Two-axis diffractometer rotation for bypassing the resolution limit of the zone plate optics

There are two types of rotations in our prototype platform. The out-of-plane rotation is used to change the incident angle of the sample, enabling resolution enhancement in one direction. The in-plane rotation is used to change the orientation of the sample. By combining in-plane rotation with out-of-plane rotation, we can improve the resolution along any pre-defined direction. In this section, we will first discuss the use of out-of-plane rotation for resolution enhancement and then discuss the combination of both out-of-plane and in-plane rotations.

For out-of-plane rotation, we rotated the stage 1 from -9 degrees to +9 degrees (with 1 degree per step) and captured the sample images accordingly. As such, the captured 19 raw images correspond to 19 different incident angles. Based on these captured raw images, we used the FP algorithm to recover the high-resolution image of the sample. Different from the original FP approach, we excluded the captured image with normal incidence in the recovery process, due to its strong zero-order diffraction.

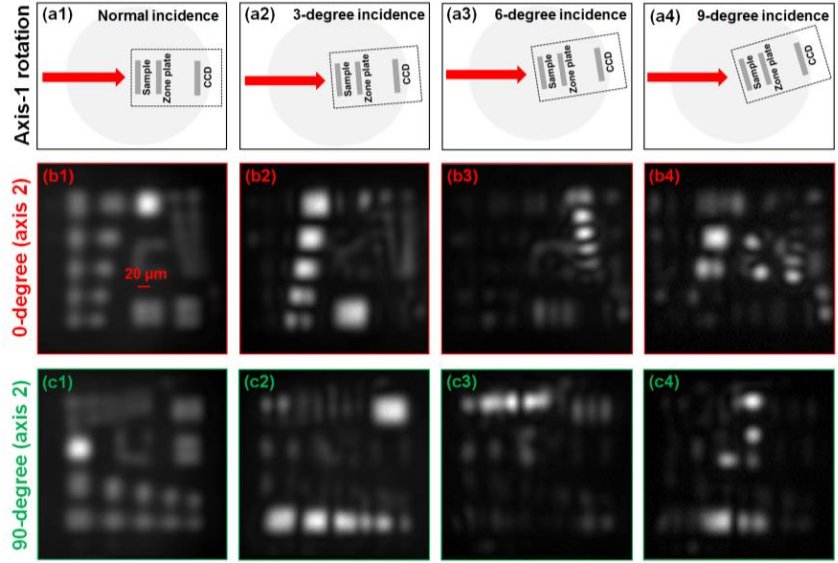

Fig. 2 Raw images of a resolution target by rotating the setup to different out-of-plane and in-plane orientations. (a) Schematic of the setup (only showing stage 1). (b1)-(b4) Raw images with incident angles ranging from 0 degree to 9 degrees (axis 1). The in-plane rotation angle is set to 0 degree (axis 2). (c1)-(c4) The in-plane rotation angle is set to 90 degrees (axis 2). By comparing Fig. 2(b1) and 2(c1), one can see that the orientation of the sample is rotated by 90 degree with axis-2 in-plane rotation.

The recovery algorithm starts with a high-resolution estimate of the sample in the Fourier domain. Next, this sample estimate is sequentially updated with the raw images corresponding to different incident angles. For each updating step, we select a small sub-region of the spectrum estimate, corresponding to one incident angle, and apply Fourier transformation to generate a new low-resolution target image. We then replace the target image's amplitude component with the measurement to form an updated, low-resolution target image. This image is then used to update its corresponding sub-region of the sample spectrum in the Fourier domain. The replace-and-update sequence is repeated for all captured images, and we iterate through the above process several times until the solution converges [5]. The final achievable resolution of the reported platform is determined by the largest out-of-plane rotation angle, not the NA of the employed binary zone plate lens.

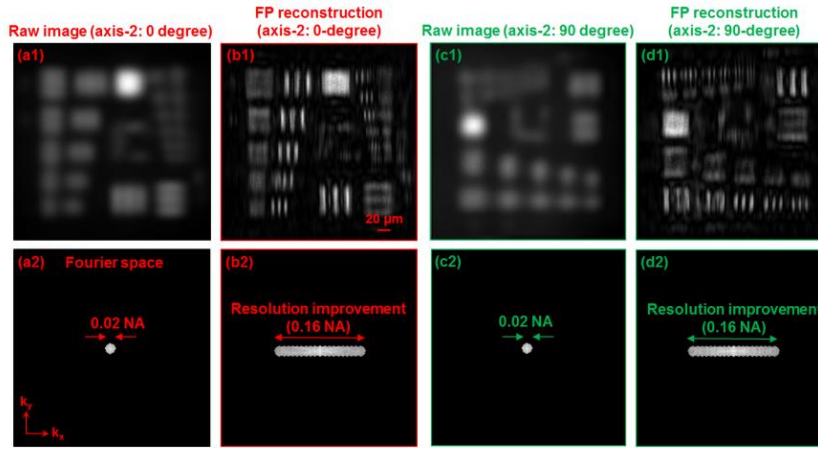

Fig. 3 FP raw images and reconstructions. (a), (c) Raw images corresponding to normal incidence (0.02 NA). (b), (d) The FP reconstructions using 19 raw images. The final synthetic NA is 0.16, along the tangent direction of the rotation stage.

Figure 3 shows the captured raw images and the FP reconstructions following the above procedures. Fig. 3(a1) and 3(c1) are raw images with 0-degree and 90-degree sample orientations, and their corresponding Fourier spectrum are shown in Fig. 3(a2) and 3(c2). Fig. 3(b1) and 3(d1) are FP reconstructions using 19 raw images. The synthetic NA of the final reconstruction is 0.16 along the tangent direction of the rotation stage 1, as shown in Fig. 3(b2) and (d2). From FP reconstructions, we can resolve the details in group 7, element 6 (linewidth of 2.19 $\mu$m), in a good agreement with the 0.16 synthetic NA.

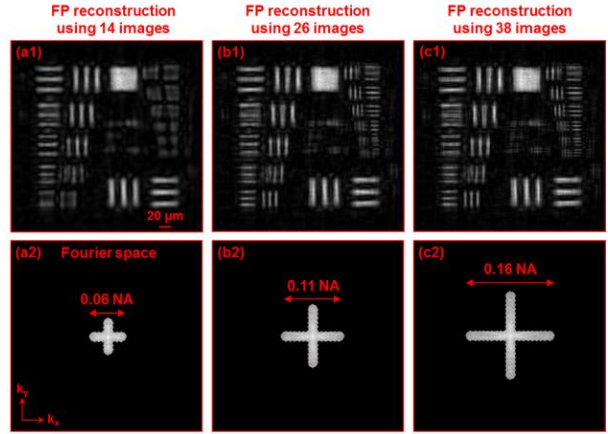

Fig. 4 FP reconstructions of a resolution target by combining both out-of-plane and in-plane sample rotations. (a) FP reconstruction with 7 raw images in x direction and 7 raw images in y direction. (b) FP reconstruction with 13 raw images in x direction and 13 raw images in y direction. (c) FP reconstruction with 19 raw images in x direction and 19 raw images in y direction.

If we only perform out-of-plane rotation of the imaging platform, we can only improve resolution along one direction. It is important to combine both out-of-plane and in-plane rotations for resolution improvement over the entire 2D Fourier space. In Fig. 3, we captured two set of raw images: one set of 19 raw images for 0-degree sample orientation (Fig. 3(a)) and one set of 19 raw images for 90-degree sample orientation (Fig. 3(c)). To combine these two set of images in the Fourier domain, we first rotated the second set of images by 90 degrees and align them with the first set measurements (using phase correlation [32]). In this case, we get 38 images of the sample, corresponding to incident angles along two different orientations (i.e., x and y directions). We then combined these two set of raw images in the Fourier domain using the FP algorithm. In Fig. 4, we demonstrate the use of these two set of images for resolution improvement along x and y directions. As is shown in Fig. 4(a), 7 images in x direction and 7 images in y direction are used for FP reconstruction. Figure 4(b) and Fig. 4(c) demonstrate the FP reconstruction results using 13+13 and 19+19 images, respectively. From Fig. 4(a2)-(c2), one can see that the improvement of resolution along two different directions has been achieved using this strategy.

Finally, we also tested our imaging platform using a titanium dioxide sample (depositing $TiO_2$ micro-particles on a glass slide). Fig. 5(a1) shows one raw image of the sample and Fig. 5(a2) shows its Fourier spectrum with cut-off frequency given by a 0.02 NA. Figure 5(b1) and (c1) show the FP reconstructions with resolution improvement along 2 and 4 directions, respectively. The corresponding Fourier spectrum are shown in Fig. 5(b2)-(c2). Fig. 5(d) shows reference image taken with 10X, 0.25 NA microscope objective lens. The resolution improvement has been highlighted by the dash-circles in Fig. 5(a1)-(c1) and Fig. 5(d).

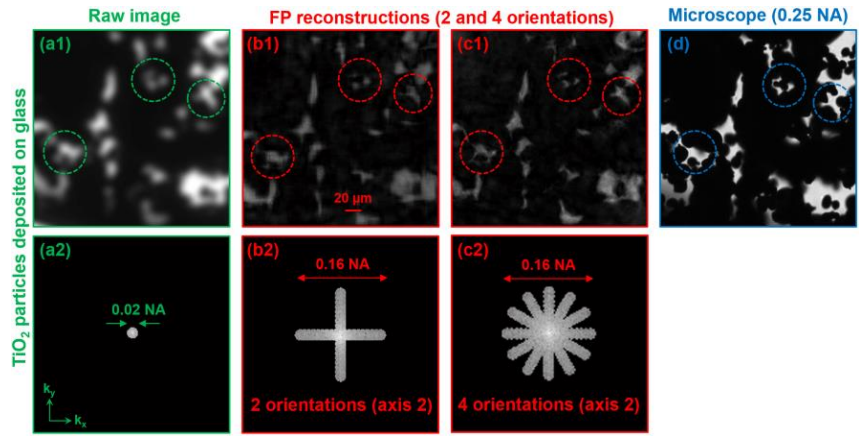

Fig. 5 FP reconstructions of a titanium dioxide sample by combining both out-of-plane and in-plane sample rotations. (a1) Raw image of the sample. (b1) FP reconstruction with resolution improvement along 2 orientations (19*2 raw images). (c1) FP reconstruction with resolution improvement along 4 orientations (19*4 raw images). (a2)-(c2) The corresponding Fourier spectrum of (a1)-(c1). (d) Reference image taken with a 10X, 0.25 NA microscope objective lens.

## 4. Conclusion

In summary, we demonstrated the use of a diffractometer setup and the FP algorithm for bypassing the resolution limit of the diffractive zone plate optics. In particular, we combined both in-plane and out-of-plane sample rotations to improve the image resolution along multiple directions in the Fourier domain. The final achievable resolution of the reported approach is not determined by the smallest width structures of the employed binary zone plate; instead, it is determined by the maximum angle of the out-of-plane rotation. In this experiment, we demonstrated 8-fold resolution improvement using both a resolution target and a titanium dioxide sample. The reported approach may be able to address the fabrication challenge of diffractive optical elements and open up new avenues for microscopy imaging with extreme wavelengths. One of our ongoing efforts is to model the high-order diffraction of the zone plate optics in the image formation and the Fourier ptychographic reconstruction process. By doing so, we aim to recover a high-resolution sample image with higher image quality.

## Funding

National Natural Science Foundation of China (NSFC) (61975254); '111 Project' (B17035); China Scholarship Council (CSC) (201806960045).

## Declaration of Competing Interest

The authors declare no any competing interests.